\documentclass{aa}

\usepackage{txfonts}
\usepackage{graphicx}
\usepackage{rotating}
\usepackage{natbib}
\usepackage{lscape}
\usepackage{epsfig}
\usepackage{hyperref}

\def\ut#1{\mathop{\vtop{\ialign{##\crcr
     $\hfil\displaystyle{#1}\hfil$\crcr\noalign
     {\kern1pt\nointerlineskip}\hbox{$\hfil\sim\hfil$}\crcr
     \noalign{\kern1pt}}}}}

\def\undersymbol#1#2{\mathop{\vtop{\ialign{##\crcr
     $\hfil\displaystyle{#2}\hfil$\crcr\noalign
     {\kern1pt\nointerlineskip}\hbox{$\hfil#1\hfil$}\crcr
     \noalign{\kern1pt}}}}}
\def\arcsec{^{\prime\prime}}
\def\arcmin{^{\prime}}
\def\degr{^0}
\def\hour{^{\rm h}}
\def\minute{^{\rm m}}
\def\second{^{\rm s}}

\newcommand{\brg}{Br$\gamma$}
\newcommand{\pab}{Pa$\beta$}

\newcommand{\hi}{\ion{H}{i}}

\newcommand{\hei}{\ion{He}{i}}

\newcommand{\um}{$\mu$m}

\newcommand{\lsun}{L$_{\odot}$}
\newcommand{\msun}{M$_{\odot}$}
\newcommand{\msunyr}{M$_{\odot}$\,yr$^{-1}$}

\newcommand{\lstar}{$L_{\mathrm{*}}$}
\newcommand{\mstar}{$M_{\mathrm{*}}$}
\newcommand{\teff}{$T_\mathrm{eff}$}

\begin{document}

\title{First X-ray detection of the young variable V1180 Cas}

\author{S. Antoniucci\inst{1}, A. A. Nucita\inst{2,3}, T. Giannini\inst{1}, D. Lorenzetti\inst{1}, B. Stelzer\inst{4}, D. Gerardi\inst{2}, S. Delle Rose\inst{2}, A. Di Paola\inst{1}, M. Giordano\inst{2,3}, L. Manni\inst{2,3}, F. Strafella\inst{2}}

\institute{
INAF-Osservatorio Astronomico di Roma, Via Frascati 33, I-00040, Monte Porzio Catone, Italy 
\and
Department of Mathematics and Physics {\it E. De Giorgi}, University of Salento, Via per Arnesano, CP 193, I-73100,
Lecce, Italy 
\and 
INFN, Sez. di Lecce, via per Arnesano, CP 193, I-73100, Lecce, Italy
\and
INAF-Osservatorio Astronomico di Palermo, Piazza del Parlamento 1, 90134, Palermo, Italy
}

\offprints{Corresponding author, \email{simone.antoniucci@oa-roma.inaf.it}}
\date{Received date / Accepted date}

  \abstract
   {V1180 Cas is a young variable that has shown strong photometric fluctuations ($\Delta I \sim 6$ mag) in the recent past, which have been attributed to events of enhanced accretion. The source has entered a new high-brightness state in September 2013, which we have previously analysed through optical and near-infrared spectroscopy.
   }
   {To investigate the current active phase of V1180 Cas, we performed observations with the Chandra satellite aimed at studying the X-ray emission from the object and its connection to accretion episodes.}
   {Chandra observations were performed in early August 2014. Complementary $JHK$ photometry and $J$-band spectroscopy were taken at our Campo Imperatore facility to relate the X-ray and near-infrared emission from the target.}
   {We observe a peak of X-ray emission at the nominal position of V1180 Cas and estimate that the confidence level of the detection is about $3\sigma$. The observed signal corresponds to an X-ray luminosity $L_X$(0.5-7 kev) in the range $0.8 \div 2.2$ $\times 10^{30}$ erg s$^{-1}$.
   Based on the relatively short duration of the dim states in the light curve and on stellar luminosity considerations, we explored the possibility that the brightness minima of V1180 Cas are driven by extinction variations. From the analysis of the spectral energy distribution of the high state we infer a stellar luminosity of 0.8-0.9 \lsun\ and find that the derived $L_X$ is comparable to the average X-ray luminosity values observed in T Tauri objects.
  Moreover, the X-ray luminosity appears to be lower than the X-ray emission levels around 5$\times10^{30} \div 1\times 10^{31}$ erg s$^{-1}$ detected at outbursts in similar low-mass objects.}
   {Our analysis suggests that at least part of the photometric fluctuations of V1180 Cas might be extinction effects rather than the result of accretion excess emission. However, because the source displays spectral features indicative of active accretion, we speculate that its photometric variations might be the result of a combination of accretion-induced and extinction-driven effects, as suggested for other young variables, such as V1184 Tau and V2492 Cyg.}


   \authorrunning{Antoniucci et al.}
   \titlerunning{X-ray detection of V1180 Cas}
   \maketitle

\section{Introduction}

A small (fraction of magnitude) and irregular photometric variability attributable to fluctuations in the mass accretion rate is a typical feature 
of all the classical T Tauri stars. However, several young sources display powerful outbursts that generate much larger brightness variations (up to 4-5 mag).
These objects are usually classified into two major classes: (1) FUors (Hartmann \& Kenyon 1985) characterized by bursts of long duration (tens of years) with accretion rates in the range of 10$^{-4}$-10$^{-5}$ M$_{\odot}$~yr$^{-1}$ and spectra dominated by absorption lines;
(2) EXors (Herbig 1989) with shorter outbursts (months--one year) with a recurrence time of months to years, showing accretion rates of the order of 10$^{-6}$-10$^{-7}$ M$_{\odot}$~yr$^{-1}$ and characterized by emission line spectra (e.g. Herbig 2008; Lorenzetti et al. 2009, 2012; Kospal et al. 2011; Sicilia-Aguilar et al. 2012; Antoniucci et al 2014a).

In particular, the nature of EXors, which spectroscopically resemble T Tauri stars in quiescence (e.g. Lorenzetti et al. 2007; Sipos et al. 2009; Antoniucci et al. 2013), is still very uncertain.
Although there is general consensus about the nature of the outbursts (i.e. events of enhanced magnetospheric accretion from the circumstellar disk), the physical mechanism that regulates the outbursts and the way they are triggered have not been clarified yet.
Aiming at shedding light on this subject, the number of optical and near-IR studies of EXors have rapidly increased in the past decade (see for example the review by Audard et al. 2014). 
Still, little is known about the X-ray properties of these sources, in particular whether 
and how much the X-ray emission from the corona of the star is affected by the accretion outburst events. 

For instance, evidence of strong X-ray emission enhancements of more than one order of magnitude during the outburst phase have been found in the cases of V1647 Ori (Kastner et al. 2004) and HBC 722 (Liebhart et al. 2014), while only modest X-ray variations have been reported for the bursts of V1118 Ori (Audard et al. 2005, 2010). No significant X-ray flux changes were detected also for Z CMa (Stelzer et al. 2009), although in this case the optical-infrared brightness increase would be primarily ascribed to variations of the circumstellar dust distribution and not to an accretion outburst (Szeifert et al. 2010).

A case of interest is that of V1180 Cas, a young variable associated with the dark cloud Lynds 1340 that has displayed strong  
photometric variations ($\Delta I \sim 6$ mag) in the recent past, which have been attributed to EXor-like mass accretion events (Kun et al. 2011).
Since September 2013, V1180 Cas is in a new high-brightness state which we have investigated through optical and near-infrared spectroscopy in a previous paper (Antoniucci et al. 2014a), confirming the presence of typical emission features that indicate accretion and mass ejection (e.g. Antoniucci et al. 2011).

To study the X-ray emission from the object during its current outburst phase, we obtained a Director's Discretionary Time (DDT) Chandra run.  
The satellite observations were coordinated with observations in the near-IR from our facilities at the Campo Imperatore Observatory, so as to have simultaneous multi-band datasets for the analysis of the source.

In this paper we report the first X-ray detection of V1180 Cas and discuss the implications of the measured X-ray emission level for understanding the nature of its current high-luminosity phase.

\section{Chandra observations}
\label{sect:chandra}

The DDT observation (obs. id. 16631) was performed on 2014 Aug
3 (JD 2456873) for a total exposure time of $\simeq 23$ ks.
We reprocessed the event files following the standard procedures
for analysis of the Chandra data using the CIAO tool suite
(version 4.6). The background level during the observation was
nominal apart from a spatial region of the original bias map for
CCD \#4 of ACIS-S, which showed unusually high values. This
did not affected our target, which was positioned on ACIS-S 3. 

In Figure 1, we show a $30\times 30$ arcsec portion of the Chandra $0.5-7$ keV band 
image centred on the nominal position of V1180 Cas 
(J2000, $RA=$ $02\hour$ $33\minute$ $01.54\second$, $DEC=$ $+72\degr$ $43\arcmin$ $26.92\arcsec$), 
which is marked with a 1-arcsec radius circle.
The CCD pixels in this area clearly show a number of counts in excess with respect to the average values of the 
surrounding pixels ($\simeq $ 0--1 counts).

We blindly searched for X-ray sources by using the {\it celldetect} tool. The method consists in sliding a square 
detect cell (with size comparable to the instrument PSF) to search for statistically 
significant counts in excess over the background. As a result, 
the signal-to-noise ratio of any detected source is computed. 
The \textit{celldetect} algorithm found a source with a signal-to-noise ratio of $\simeq 2.8$ 
at the coordinates $RA=$ $02\hour$ $33\minute$ $01.593\second$ and $DEC=$ $+72\degr$ $43\arcmin$ $26.82\arcsec$, with a statistical
error on both coordinates of $\simeq 0.2\arcsec$; this position is highly 
consistent with that of V1180 Cas (see also Sect.~\ref{sect:discussion}). 

Owing to the low number of counts detected by \textit{celldetect} ($\simeq 14$), 
we limited ourselves to estimating the broad-band (0.5-7 kev)
flux of the source following the recipe described on
http://cxc.harvard.edu/ciao/threads. In particular, we used the {\it aprates} tool to estimate the number of counts (background-corrected) of the source and 
its rate\footnote{We tested several alternative 
methods in evaluating the source net counts, with all of them resulting in consistent estimates.}. 

We extracted the source counts from a circular region 
centred on the \textit{celldetect} coordinates and with a radius of $1.5''$. The 
background was accumulated on a surrounding annulus with inner and outer 
radii of $5''$ and $10''$, respectively\footnote{This corresponds to a PSF fraction in the source aperture $\alpha$=0.928 and in the background aperture $\beta=0.01$.}.
This procedure resulted in source net counts 
of $16^{+8}_{-6}$ and a rate of $(7\pm 3)\times 10^{-4} s^{-1}$ for a livetime of 22720 s.
We then extracted the source and background spectra (as well as the associated response matrix and ancillary files) 
by using the CIAO tool {\it specextract}. 

We explored two different procedures for fitting the spectrum and derived an estimate as reliable as possible of the X-ray flux from the target.
In the first approach, we re-binned the spectrum to two counts per bin and loaded it in XSPEC (vers. 12.7.1) to fit it with an APEC model (Smith et al. 2001) using the $\chi^2$ statistic with errors remodulated through a Churazov weighting (Churazov et al. 1996). 
Given the low number of counts, we decided to limit the number of free parameters by fixing the plasma temperature to three different values of 1, 2, and 5 keV. 
The re-binned spectrum and the fit for $T=1$ keV are shown in Fig.\ref{fig:fit}, while the results of the fits are given in the upper part of Table~\ref{tab:bestfit}. For a temperature of 1 keV, the derived absorbing column density $n_H$ is around $10^{22}$ cm$^{-2}$ and the absorbed flux is 2.9 $\times 10^{-15}$ erg s$^{-1}$ cm$^{-2}$. For the higher temperatures, the absorbed flux varies within a factor two, while we find that $n_H$ and the normalization constant\footnote{We recall that in the XSPEC implementation, the normalization 
constant is $L_{39}/d_{10}^2$, where $L_{39}$ is the source luminosity in units of $10^{39}$ 
erg s$^{-1}$ and $d_{10}$ the distance to the source in units of $10$ kpc.} $N$ tend to become less constrained.

In the second method, we did not bin the data and used the Cash statistic (Cash 1979) in XSPEC. If we again fix the temperature at 1, 2, and 5 keV, we obtain no good convergence for any of the fits and can only provide upper limits for the parameters and flux (see central part of Table~\ref{tab:bestfit}). 
Alternatively, by leaving the temperature free to vary, we are able to find a best-fit solution for $kT=0.68$ keV, $n_H=1.1\times 10^{22}$ cm$^{-2}$, and $N=1.1 \times 10^{-5}$cm$^{-3}$ (see lower part of Table~\ref{tab:bestfit}) with an inferred absorbed flux of 3.1 $\times 10^{-15}$ erg s$^{-1}$ cm$^{-2}$. 
The confidence level contours of this latter fit for the $kT$ and $n_H$ pair are reported in Fig.~\ref{fig:contours} and show that the possible solutions are found in a well localized region of the parameter space. 

\begin{figure}[t]
\begin{center}
\includegraphics[width=9cm]{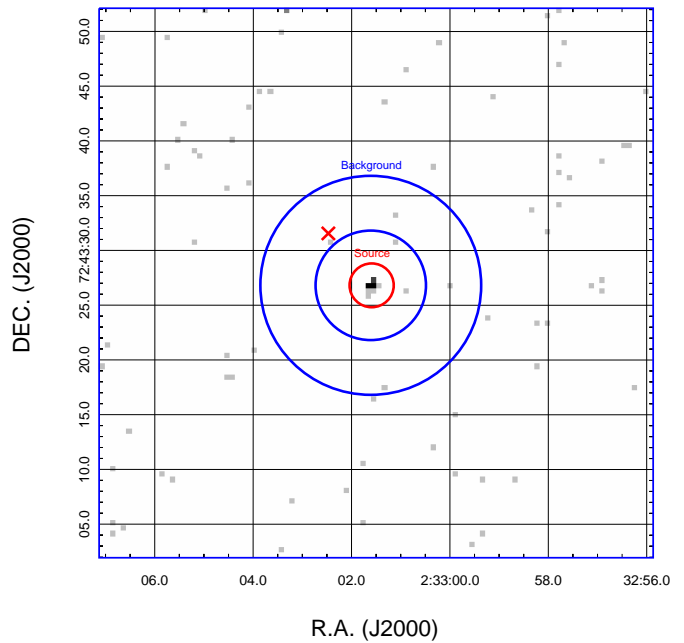}
\end{center}
\caption{\label{fig:chandra} Portion (50$\times$50 arcsec) of the Chandra image in the $0.5-7$ keV band around the nominal position of V1180 Cas. The source extraction area is marked by a red circle with a radius of 1.5$\arcsec$, while the annulus for background computation is traced in blue. The location of the infrared source V1180 Cas B is indicated by a red cross.}
\end{figure}
\begin{table*}[t]
\centering
\caption{\label{tab:bestfit} Results of the APEC spectral model fits obtained within XSPEC.}
\begin{small}
\begin{tabular}{l|ccccc}
\hline
\hline
Fit type & $n_H$/$10^{22}$ (cm$^{-2}$) &  $kT$ (keV) & $N$ ($10^{-6}$ cm$^{-3}$) &  Statistic$^{a}$/dof & $F_{0.5-7 \mathrm{keV}}^{b}$ ($10^{-15}$ erg s$^{-1}$ cm$^{-2}$) \\
\hline
                           &  1.1 ($-$0.6,1.2)  &    1                 &      7 ($-$5, 10)        &  0.8/5  &   2.9 ($-$2.4, 1.4) \\
Binned data (fixed $kT$)   &  $\le$0.94         &    2                 &      4.4 ($-$2.9, 4.9)   &  0.6/5  &   4.0 ($-$3.4, 1.9) \\
                           &  $\le$0.69         &    5                 &      3.6 ($-$1.7, 2.8)   &  0.5/5  &   5.8 ($-$3.8, 2.3) \\
\hline
                           &  0.8 ($-$0.4,0.5)  &    1                 &      $\le$7.6            &  84     &   $\le$ 3.2  \\
Unbinned data (fixed $kT$) &  $\le$0.7          &    2                 &      $\le$7.5            &  88     &   $\le$ 5.6  \\
                           &  $\le$0.4          &    5                 &      $\le$6.7            &  89     &   $\le$ 8.5  \\
\hline
Unbinned data              &  1.1 ($-$0.2,0.3)  &  0.68 ($-$0.19,0.19) &   11 ($-$6,19)           &  81     &   3.1 ($-$1.5,1.5)  \\
\hline
\end{tabular}
\tablefoot{Lower and upper errors are given in parentheses.
\tablefoottext{a}{Statistic is $\chi^2$ for binned data and $C$ for unbinned data.}
\tablefoottext{b}{Absorbed fluxes.}
}
\end{small}
\end{table*}
\begin{figure}[t]
\centering
\includegraphics[width=6.7cm,angle=-90]{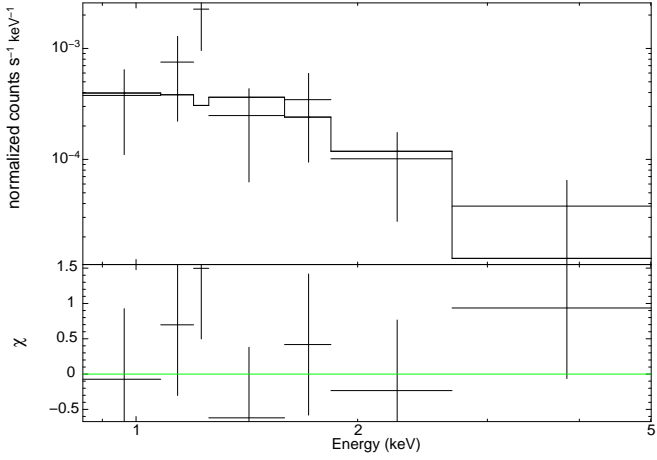}
\caption{\label{fig:fit} Observed spectra rebinned to 2 cts/bin (points with error bars) and fit (histogram) of our Chandra observation
(top plot). The shown fit is relative to the case of binned data with fixed $kT$ of 1 keV (first line of Table~\ref{tab:bestfit}). 
Residuals are shown in the bottom plot.}
\end{figure}

\begin{figure}[t]
\centering
\includegraphics[width=7cm, angle=-90]{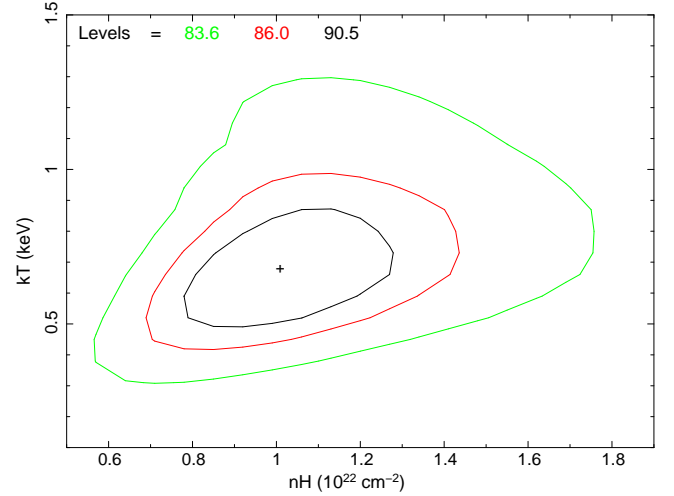}
\caption{\label{fig:contours} Confidence contours for $kT$ versus $n_H$ in the analysis of unbinned data with the C-statistic (see last row of Table~\ref{tab:bestfit}). Contour levels are indicated in the plot, while the best-fit position is marked by a cross.
}
\end{figure}


The results of the previous analysis (both binned and unbinned data) seem to suggest that we are observing emission from a plasma with a temperature $\sim$0.7 keV and a column density of $\sim$1$\times$ 10$^{22}$cm$^{-2}$ and that the total (absorbed) flux is about $3 \times 10^{-15}$ erg s$^{-1}$ cm$^{-2}$. 
By adopting the relationship $n_H \simeq 2\times 10^{21} A_V$ (mag) for standard interstellar
gas-to-dust ratio (see e.g. Güver \& Özel 2009, and references therein), we find that the inferred value of the hydrogen column density is fairly consistent with the estimated average visual extinction towards the source ($A_V \simeq 4-5$ mag, Antoniucci et al. 2014a; Kun et al. 2011), which provides further support to the results of the fit. 

The estimated absorbed flux in the 0.5-7 keV band corresponds to an unabsorbed flux of $(3.5\pm1.7)\times 10^{-14}$ erg s$^{-1}$ cm$^{-2}$, which is the value we eventually consider for the X-ray emission from V1180 Cas.

\begin{figure*}[t]
\centering
\includegraphics[width=16.0cm]{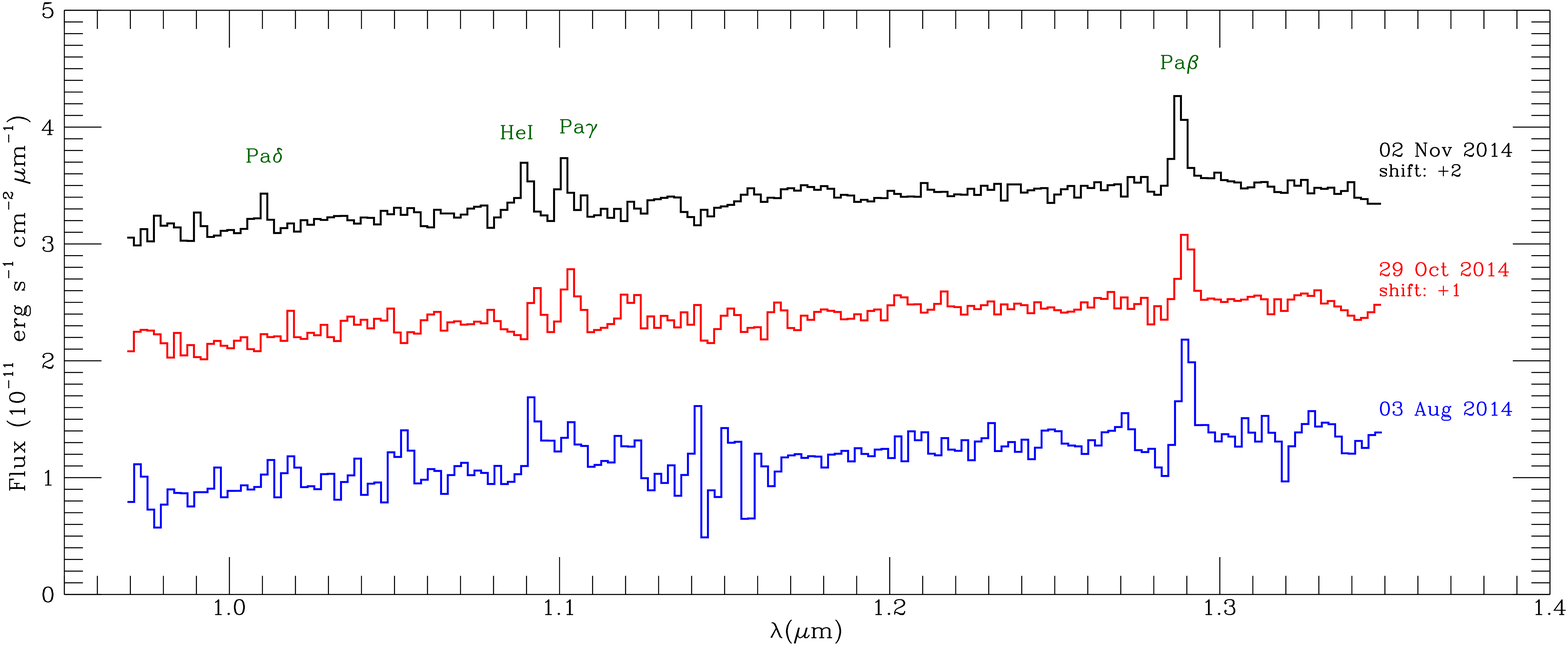}
\caption{\label{fig:specs} Near-IR $J$-band spectra of V1180 Cas taken on different dates at the Campo Imperatore Observatory, with recognisable emission features labelled. 
The upper spectra have been shifted by the indicated amount for better visualization.)} 
\end{figure*}


\begin{table}[t]
\caption{\label{tab:phot} V1180 Cas near-IR photometry.}
\begin{small}
\begin{tabular}{l|c|ccc}
\hline
\hline
Date & JD--2450000.5  &  $J$   &  $H$  &  $K$  \\
\hline
18 Nov 2013 & 6614$^a$ &  13.46 & 12.21 & 11.13 \\
03 Aug 2014 & 6873     &  13.48 & 12.21 & 11.20 \\
21 Oct 2014 & 6951     &  13.34 & 12.04 & 11.02 \\
03 Nov 2014 & 6964     &  13.34 & 12.06 & 11.02 \\
\hline
\end{tabular}
\tablefoot{
\tablefoottext{a}{Photometry taken at TNG. Typical error is 0.03 mag.}\\
}
\end{small}
\end{table}

\begin{table*}[t]
\centering
\caption{\label{tab:lines} Fluxes of detected near-IR emission lines.}
\begin{small}
\begin{tabular}{l|c|cccc}
\hline
\hline
Line ID & $\lambda$(\um) &  2013 Nov 18$^a$ & 2014 Aug 03 & 2014 Oct 29 & 2014 Nov 02 \\
\hline

Pa$\delta$       & 1.005  & 7.1  $\pm$ 0.5 & $<$ 16   & $<$ 12   & 8$\pm$3 \\
\hei             & 1.082  & 13.1 $\pm$ 0.5 & 30$\pm$6 & 16$\pm$5 & 23$\pm$3 \\
Pa$\gamma$       & 1.094  & 15.2 $\pm$ 0.5 & 19$\pm$6 & 23$\pm$5 & 17$\pm$3 \\
\pab             & 1.282  & 35.7 $\pm$ 1.0 & 40$\pm$6 & 38$\pm$5 & 37$\pm$3 \\
\brg             & 2.166  & 7.1  $\pm$ 0.3 &  9$\pm$4 & ...      & ...   \\
\hline
\end{tabular}
\tablefoot{
Fluxes are given in 10$^{-15}$ erg s$^{-1}$ cm$^{-2}$. $^a$ TNG/NICS spectroscopic observations (Antoniucci et al. 2014a).
}
\end{small}
\end{table*}

\section{Near-IR observations}
\label{sect:nir}
Infrared observations were performed on 2014 Aug 3 (the same day as Chandra observations) with SWIRCAM (D'Alessio et al. 2000), which is mounted on the 1.1m AZT-24 telescope of the Campo Imperatore Observatory (L'Aquila, Italy). 
Photometric images in the $JHK$ near-IR broad-band filters were acquired by dithering the telescope around the pointed position
and reduced by using standard procedures for bad-pixel removal, flat-fielding, and sky subtraction.

In addition, low-resolution ($\mathcal{R}$ $\sim$ 250) spectra were acquired with the G$_{blue}$ and G$_{red}$ infrared grisms  
that cover the $ZJ$ (0.83–1.34 \um) and $HK$ (1.44–2.35 \um) wavelength segments, making use of the standard
ABB$^\prime$A$^\prime$ technique with a total integration time of about 30 minutes.
The raw spectral frames were flat-fielded, sky-subtracted, and corrected for
the optical distortion in both the spatial and spectral directions.
Telluric features were removed by dividing the extracted spectra
by that of a normalized standard star, after removing any intrinsic spectral features. 
Wavelength calibration was achieved from sky OH emission lines, while
flux calibration was derived from the photometric data.
SWIRCAM observations were repeated on 2014 Oct 21 and Nov 3 ($JHK$ photometry) and on Oct 29 and Nov 2 (spectroscopy, only in the $ZJ$ bands). 

Derived $JHK$ magnitudes are reported in Table~\ref{tab:phot}, while the fluxes of detected emission lines are given in Table~\ref{tab:lines}. Values measured on 2013 Nov 18 with NICS (Antoniucci et al. 2014a) at the Telescopio Nazionale Galileo (TNG) are also reported in the table for comparison.
The three G$_{blue}$ spectra acquired on different dates are displayed in Fig.~\ref{fig:specs}.

Our photometric data show that the near-IR magnitudes and colours of V1180 Cas have remained substantially unaltered in comparison with observations that we performed in 2013 Nov, thus indicating that the high-brightness state of the source is fairly stable on time scales of several months. Our measurements indeed suggest that the current state might be similar to the previous long-lasting high-luminosity period observed by Kun et al. (2011) between 2007 and 2010.

The main $J$-band emission lines (\hi\ and \hei) that were detected in our TNG/NICS spectrum of 2014 November are still recognizable in all the spectra (see Fig.~\ref{fig:specs}), despite the much lower sensitivity of the Campo Imperatore instrumentation. These new spectra are characterized by different signal-to-noise ratios, with the spectrum taken in August showing strong residuals from telluric absorptions.
The hydrogen lines do not show significant flux variations from values reported in Antoniucci et al. (2014a) and look fairly stable at the different dates, whereas
the 1.08~\um\ helium line, which is a known tracer of stellar winds (e.g. Edwards et al., 2006; Ray et al., 2007), shows evidence of some degree of variability.
The similarity of the \hi\ emission line fluxes indicates that the mass accretion rate has not substantially changed from the value of 3$\times$ 10$^{-8}$ \msunyr\ that was estimated in 2013 November by Antoniucci et al. (2014a) from several emission features (including those observed in the $J$ band).

\section{Discussion}
\label{sect:discussion}

\subsection{X-ray source position}
\label{sect:disc1}

To our knowledge, no previous detections of V1180 Cas have been reported in the literature; in particular, we found no ROSAT (all-sky survey) or XMM-Newton object (3XMM-DR4 catalogue, Watson et al., in preparation).
As stated in Sect.~\ref{sect:chandra}, the coordinates of the detected X-ray source are highly consistent with the nominal 
position of V1180 Cas. Given the spatial resolution of the Chandra observation, we can rule out that the X emission is 
associated with the infrared object V1180 Cas B (Kun et al. 2011), which is located at an angular distance of about
6$\arcsec$ NE of V1180 Cas. For similar reasons, the observed X-ray emission peak cannot be related to the known jet shocks that are 
observed farther north of both sources, which are likely driven by V1180 Cas B (Antoniucci et al. 2014a).

\begin{table}[]
\caption{\label{tab:sed} Magnitudes of the high state of V 1180 Cas.}
\begin{small}
\begin{tabular}{lccc}
\hline
\hline
Filter/band & mag & Epoch & Reference \\
\hline
 B        &  20.4          & 2014 Feb & 1 \\
 V        &  18.6          & 2014 Feb & 1 \\
 R        &  17.0          & 2014 Feb & 1 \\
 I        &  15.4          & 2014 Feb & 1 \\
 J        &  13.4          & 2013 Nov - 2014 Aug & 1, this paper \\
 H        &  12.1          & 2013 Nov - 2014 Aug & 1, this paper \\
 K        &  11.1          & 2013 Nov - 2014 Aug & 1, this paper \\
 WISE 3.4\um   &    9.821  & 2010 Jun 2010 & 2 \\
 WISE 4.6\um   &    8.422  & 2010 Jun 2010 & 2 \\
 WISE 11.6\um  &    5.547  & 2010 Feb 2010 & 2 \\
 WISE 22.1\um  &    3.590  & 2010 Feb 2010 & 2 \\
 IRAC 3.6\um   &    9.58   & 2009 Mar & 3 \\
 IRAC 4.5\um   &    8.58   & 2009 Mar & 3 \\
 IRAC 5.8\um   &    7.73   & 2009 Mar & 3 \\
 IRAC 8.0\um   &    6.67   & 2009 Mar & 3 \\
 MIPS 24\um    &    3.64   & 2009 Mar & 3 \\
 \hline
\end{tabular}
\tablefoot{
References: 1: Antoniucci et al. 2014a; 2: ALLWISE data release (Cutri et al. 2013); 3: Kun et al. 2011.
}
\end{small}
\end{table}

\subsection{X-ray luminosity and nature of the high-brightness state}
\label{sect:disc2}

From our Chandra measurements and assuming a distance of 600 pc for the target (Kun et al., 1994), we can derive an X-ray luminosity $L_{X}$~(0.5-7 kev) 
in the range $0.8 \div 2.2$ $\times 10^{30}$ erg s$^{-1}$ including the uncertainties.
To compare $L_{X}$ to the stellar luminosity (\lstar) of the source, we can refer to the stellar parameters estimated by Kun et al. (2011). These authors derived \teff\ = 4060 K and \lstar\ = 0.07 \lsun, on the basis of the flux ratios measured in the optical spectrum and from the $I$ and $J$ magnitudes at minimum brightness, i.e. the phase where we can observe the actual emission from the stellar photosphere (in the accretion outbursts scenario).
However, Kun et al. (2011) warn that a stellar luminosity of only 0.07 \lsun\ is clearly too low, since it would place the source close to the main sequence. They therefore argue that this result is due to scattered light dominating the emission at minimum brightness, which is indicative of an underestimation of the circumstellar extinction. Indeed, adopting such a low value of \lstar, we derive a fairly high Log $L_X/L_{*}$ between $-2.5$ and $-2.1$.

On the other hand, the considerations by Kun et al. (2011) and the relatively short duration (if compared to typical quiescence periods of other eruptive variables) of the recent dims of V1180 Cas between 2003--2006 and 2010--2011 (see light curve in Kun et al., 2011) might also indicate that the brightness minima in the light curve are caused by a temporary increment of the extinction value along the line of sight, analogously to UXor objects (e.g. Shenavrin et al. 2012). 
In this alternative scenario, the current high state would roughly represent the normal state of the object, with emission at short wavelengths dominated by photons from the photosphere of the star.

\begin{figure}[t!]
\centering
\includegraphics[width=9.5cm]{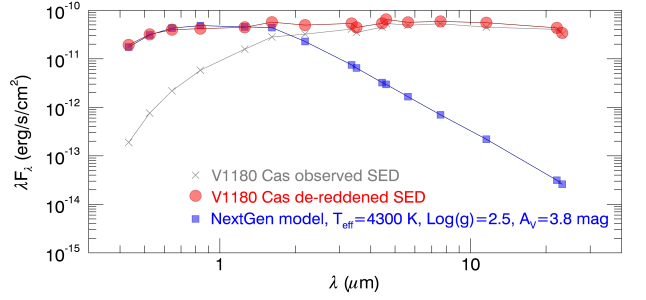}
\caption{\label{fig:sed} High state SED of V1180 Cas constructed with the photometry listed in Table~\ref{tab:sed}. The de-reddened SED is shown in red, whereas the observed points are plotted in grey. The best-fit NextGen stellar model obtained with the VOSA is traced in blue (model parameters and fitted extinction are given in the legend.)} 
\end{figure}

To investigate this aspect, we collected optical-to-mid-infrared photometry from this and previous articles, which is reported in Table~\ref{tab:sed}. All these measurements refer to observations taken while the object was in a high state (see the $I$-band light curve of Kun et al. 2011). Under the assumption that the target behaviour is the same during each high-brightness phase, we are therefore able to construct the spectral energy distribution (SED) of the source in its high state, which is shown in Fig.~\ref{fig:sed}. Long-wavelength points clearly indicate an infrared excess that is associated with the circumstellar disc. The presence of such disc was also signalled by the presence of CO 2.3\um\ bands in emission in the spectra of Antoniucci et al. (2014a). 

To derive the stellar luminosity of the source, we fitted the SED points shortwards of the infrared excess (which we assume to start from the $K$ band) using the Virtual Observatory SED Analyzer (VOSA; Bayo et al. 2008) with a NextGen grid of models (Allard et al. 2012) and considering a 10\% uncertainty on the fluxes.
Adopting the constraints log $g$ $\geq 2$ and $A_V$ = 5$\pm$2 mag, we obtain a set of five best-fit models with \teff\ in the range 4200-4400 K and \lstar\ $\sim$ 0.8-0.9 \lsun\, and visual extinction estimates between 3.6 and 4.0 mag. The best-fit model with a $\chi^2$ around 10 is plotted in Fig.~\ref{fig:sed}. We notice that the estimated \teff\ is not too different from the value of 4060 K estimated by Kun et al. (2011) from optical spectral features. 
This stellar luminosity would imply Log $L_X/L_{*}$ in the range -3.6 $\div$ -3.2 for the source and mass in the range 0.6--1.0 \msun, based on Siess et al. (2000) evolutionary models. 
We point out that an $L_X$ around $10^{30}$ erg s$^{-1}$, like the one we obtain for V1180 Cas, is in agreement with the X-ray luminosities observed in classical T Tauri stars with \mstar$\sim$1\msun\ and \lstar$\sim$1\lsun\ (see statistical correlations by Telleschi et al. 2007).

Therefore, in the scenario where the high state is the one where the emission from the stellar photosphere dominates, the X-ray flux from V1180 Cas seems to be comparable to standard emission from T Tauri stars. On the other hand, if the stellar luminosity of the object is significantly lower (as expected if the current optical-near IR luminosity is affected by an ongoing outburst), the detected X-ray flux would suggest that the X-ray emission in this phase is somewhat enhanced, because an emission level of the order of $10^{30}$ erg s$^{-1}$ would be more than one order of magnitude greater than standard X-ray fluxes from T Tauri stars with \lstar$\sim$0.1 \lsun. 

Since no X-ray measurement for the low-brightness state of V1180 Cas is available, we do not have any reference to understand whether the X-ray emission from the source is currently increased and whether it varies according to optical and infrared fluctuations. We can, however, take the absolute value of $L_X$  into account to understand if the  X-ray flux is comparable to that of similar low-mass objects in outburst.
Liebhart et al. (2014) inferred $L_X \sim 4 \times 10^{30}$ for HBC 722 (a less massive, 0.5 \msun\ source), while Kastner et al. (2004) and Teets et al. (2011) derived a luminosity of the order of $10^{31}$ erg s$^{-1}$ for V1647 Ori (0.8 \msun\ object). In both cases there was also a significant X-ray flux increase at outburst with respect to the quiescent phase (see for example the $\sim3 \times 10^{29}$ erg s$^{-1}$ in quiescence reported for V1647 by Kastner et al. (2004), which has been directly connected to the accretion event.
That these outburst values are greater than the flux we derive for V1180 Cas may provide additional hints at a scenario in which the current phase of the source is not dominated by an ongoing outburst, always assuming that we are not underestimating $n_H$ and that the presumably enhanced X-ray emission regime is not decaying faster than the optical-infrared variations.

We notice in addition that hydrogen column densities that are higher than those expected on the basis of the estimated optical extinction and standard gas-to-dust ratio have recently been derived by Liebhart et al. (2014) for the outburst of HBC 722, indicating the presence of large amounts of dust-free gas in the circumstellar environment during the accretion event. On this basis, that for V1180 Cas we find a value of $n_H$ that is consistent with the estimated $A_V$ (Sect. \ref{sect:chandra}) may be considered as a further support to the scenario where no major accretion event is driving the brightness variations. 

In conjunction with the long duration of the recent high states of V1180 Cas, some of the previous considerations suggest that the photometric behaviour of the target might be at least in part driven by recurrent obscuration effects. 
In any case, since optical and near-IR spectroscopic analyses have clearly revealed typical features indicative of ongoing accretion activity, such as veiling, CO bands in emission, and \hi\ emission lines (Kun et al. 2011; Antoniucci et al. 2014a, 2014b), we speculate that the photometric variations of V1180 Cas are the result of a combination of accretion-related and extinction-driven effects, as proposed in the cases of the pre-main sequence variables V2492 Cyg (Hillenbrand et al. 2013) and V1184 Tau (Grinin et al. 2009), whose light curves present several similarities with that of V1180 Cas.
For the brightness dims of V2492 Cyg, Hillenbrand et al. (2013) suggest semi-periodic occultations of the innermost regions by an orbiting dusty clump or disk warp. Similarly, Grinin et al. (2009) hypothesize that the faint states of V1184 Tau are due to obscuration of the central star region by the puffed-up inner rim of the disk, whose height would increase during higher accretion rate phases.

\section{Conclusions}
By using the Chandra satellite, we obtained the first X-ray detection of V1180 Cas, a young variable that has shown strong photometric fluctuations ($\Delta I \sim 6$ mag) in the course of last 15 years and has entered a new high-brightness state in September 2013. 
Because of the low number of counts detected, we limited our analysis to computing an estimate of the broad-band flux $L_X$ between 0.5 and 7 keV, by fitting a simple thermal model to the data.

Simultaneous and complementary $JHK$ photometry and $J$-band spectroscopy were carried out using the Campo Imperatore facility to relate the X-ray and near-infrared emission from the target.
Our near-infrared datasets show that the photometry and emission line flux of V1180 Cas have not significantly changed from the previous measurements that we performed in November 2013, thus confirming that the current state of the object is fairly stable.

From the observed Chandra signal, we derived an estimate of the X-ray luminosity $L_X$(0.5-7 kev) in the range $0.8 \div 2.2$ $\times 10^{30}$ erg s$^{-1}$.
Based on the relatively short duration of the dim states in the light curve and on the fact that the stellar luminosity inferred from the colours at minimum brightness appears too low, we considered the possibility that the brightness minima of V1180 Cas are driven by extinction variations. By constructing the SED of the high state, we therefore inferred a stellar luminosity of 0.8-0.9 \lsun\ and found that the derived $L_X$ is actually comparable to the average X-ray luminosity values observed in T Tauri objects.
In addition, we note that the measured X-ray flux is somewhat lower than the enhanced X-ray emission levels of the order of 5$\times10^{30} \div 1\times 10^{31}$ erg s$^{-1}$ detected at outbursts in similar low-mass objects such as HBC 722 and V1647 Ori.

These results provide hints in favour of the scenario in which part of the photometric fluctuations of V1180 Cas are extinction effects rather than the result of accretion-related excess emission. However, since the source displays spectral features indicative of active accretion, its photometric behaviour might be explained by a combination of accretion-induced and extinction-driven effects, as suggested in the cases of the young variables V2492 Cyg and V1184 Tau.

New and deeper X-ray observations of V1180 Cas (especially during its low-brightness state) are encouraged to check for fluctuations in this band and provide support to the proposed scenario.

\subsection*{Acknowledgements}
\begin{footnotesize}
The authors are very grateful to the referee, M. G\"udel, for his insightful comments and suggestions that helped us to improve the quality of the paper.

\noindent This publication makes use of VOSA, developed under the Spanish Virtual Observatory project supported by the Spanish MICINN through grant AyA2011-24052.

\noindent SA acknowledges support from the T-REX-project, the INAF (Istituto Nazionale di Astrofisica) national project aimed at maximizing the participation of astrophysicists and Italian industries
in the E-ELT (European Extremely Large Telescope). The T-REX project has been approved and funded by the Italian Ministry for Research and University (MIUR) in the
framework of “Progetti Premiali 2011” and then “Progetti Premiali 2012”.
\end{footnotesize}


\end{document}